# M Dwarfs, Microlensing, and the Mass Budget of the Galaxy

John N. Bahcall[1], Chris Flynn[1] Andrew Gould[2], and Sofia Kirhakos[1]

E-mail jnb@sns.ias.edu

## Abstract

We show that faint red stars do not contribute significantly to the mass budget of the Galaxy or to microlensing statistics. Our results are obtained by analyzing two long exposures of a high-latitude field taken with the Wide Field Camera on the newly repaired *Hubble Space Telescope*. Stars are easily distinguished from galaxies essentially to the limiting magnitudes of the images. We find five stars with $2.0 < V - I < 3.0$ and $I < 25.3$ and no stars with $V - I > 3.0$. Therefore, main-sequence stars with $M_I > 10$ that are above the hydrogen-burning limit in the dark halo or the spheroid contribute $< 6\%$ of the unseen matter. Faint red disk stars, M-dwarfs, contribute at most 15% to the mass of the disk. We parameterize the faint end of the cumulative distribution of stars, $\Phi$, as a function of luminosity $L_V$, $d\Phi/d\ln L_V \propto L_V^{-\gamma}$. For spheroid stars, $\gamma < 0.32$ over the range $6 < M_V < 17$, with 98% confidence. The disk luminosity function falls, $\gamma < 0$, for $15 \lesssim M_V \lesssim 19$. Faint red stars in the disk or thick disk, and stars with $M_V < 16$ in the spheroid contribute $\tau < 10^{-8}$ to the optical depth to microlensing toward the Large Magellanic Cloud.

Subject Headings: dark matter – gravitational lensing – stars: low mass – stars: luminosity function

1 Building E, Institute For Advanced Study, Princeton, NJ 08540
2 Dept of Astronomy, Ohio State University, Columbus, OH 43210

1. Introduction

Historically, the direct detection of faint stars has been limited by the difficulty in distinguishing with ground-based observations faint images of stars from galaxies, which are much more numerous at faint magnitudes ($V \gtrsim 20$). Here we present the results of a search for faint red Galactic stars, as originally suggested by Bahcall, Guhathakurta, & Schneider (1990), in deep Wide Field Camera (WFC2) images taken with the newly repaired *Hubble Space Telescope (HST)*. The images, which greatly improve previous limits on the number of faint red stars, cover 4.4 square arcmin and total 2.00 hr in $V'$ (F606W) and 2.83 hr in $I'$ (F814W). The field contains 5 stars with $2.0 < V - I < 3.0$ and no stars with $V - I > 3.0$, down to the the magnitude limit $I < 25.3$. The field analyzed in this *Letter* ($\ell = 241$, $b = -51$), has the longest exposures among seven *HST* WFC2 fields that we are currently investigating. In a future paper, we will report our techniques for analyzing the other fields and the resulting measurements of the luminosity functions (LFs) of the disk, thick disk, and spheroid. However, we have checked one of the other fields ($\ell = 148$, $b = -57$), which has a magnitude limit $I < 24.7$, in order to make certain that the primary field is not anomalously devoid of stars. This second field contains five stars with $2.0 < V - I < 3.0$ and none with $V - I > 3.0$. It is therefore consistent with the primary field.

The resolution of these images is set by the pixel scale, $0\rlap{.}''1$. One can distinguish between stellar and extended objects down to the magnitude limit, several magnitudes deeper than all previous star-count studies. (See Bahcall 1986 for a review and e.g., Majewsky 1992; Richer & Fahlman 1992; Richstone et al. 1992, Gould, Bahcall, & Maoz 1993 for several deeper and more recent studies using a variety of techniques.) For faint red stars, we establish improved constraints on contributions to the mass of the Galaxy, the slopes of LFs, and the optical depths to microlensing. The limits on faint red stars obtained from the *HST* WFC2 star counts are stronger than either the dynamical or microlensing limits for all four Galactic components considered in this paper. We cannot put significant limits on brown dwarfs because they are too faint.

For purposes of this *Letter*, we define the faint endpoint of the main-sequence as the reddest known nearby star in $V - I$, i.e., $M_I = 14$, $V - I = 4.7$ (see Monet et al. 1992). (Fainter stellar objects are known, but these are also bluer.) We refer somewhat loosely to this color-dependent endpoint as the "hydrogen-burning limit". The actual magnitude of the faintest hydrogen-burning stars is uncertain, but is probably close to the value adopted here. For the disk, we also give results in terms of a hydrogen-burning limit of $M_I = 15$ (A. Burrows 1994, private communication).

## 2. Observations

We use 6 $V'$ and 7 $I'$ WFC2 CCD *HST* observations of a field at $l = 241.1237$, $b = -51.2286$ from the Guaranteed Time Observer's parallel observing program. These exposures ranged from 500 s to 2100 s, for a total of 7200 s in $V'$ and 10,200 s in $I'$. The individual images are aligned to better than 0.1 pixel. We combined the images by calculating their weighted and scaled means, but for each pixel we excluded in sequence the highest $\sigma$ outliers until each included pixel lay within 3 $\sigma$ of the mean of the others. This removed the cosmic-ray events efficiently. We then removed the dead pixels and hot pixels (as determined by median stacking images of 10 other fields) by a quadratic fit to the eight nearest pixels. This procedure produced combined images with background noise in electrons per second of $3.3 \times 10^{-3}$ $e$ s$^{-1}$ pixel$^{-1}$ in $I'$ and $4.7 \times 10^{-3}$ $e$ s$^{-1}$ pixel$^{-1}$ in $V'$. We constructed a catalog of stars in two stages. First, we established a list of stellar candidates on each of the resulting (800 × 800) WFC2 images in each of the two bands by searching (by eye) for objects that are not obviously extended and that satisfy one or more of the following criteria: 1) $V - I \geq 2.0$ and $I'$ flux $\geq 0.25$ $e$ s$^{-1}$, 2) $I'$ flux $\geq 0.39$ $e$ s$^{-1}$, or 3) $V'$ flux $\geq 0.54$ $e$ s$^{-1}$. Second, we subjected each candidate to a star/galaxy discrimination test. To construct this test, we examined an image of a field at low galactic latitude ($b = 7°$) with > 20 bright stars on each chip. We compared the radial profiles of all the stars from each chip, which showed that the point spread function is nearly constant over each chip. We used this superposition of radial profiles as a test template for deciding if objects are stars. All of the candidates fall clearly into 'stellar' and 'non-stellar' categories. Finally, we repeated the entire procedure replacing the 'by-eye' search for candidates with an automated search. The resulting catalog of stars was identical to the 'by-eye' search.

Any stellar object detected in $I'$ had a measurable flux in $V'$ and vice-versa. The main results of this *Letter* concern red stars; hence, the important flux limit is in $I'$. In terms of Johnson/Cousins $I$, the limiting magnitude is $I < 25.3$ for $2 < V - I < 5$. Although it is not important for this *Letter*, the limiting $V$ magnitude is $V \lesssim 25.6 + 0.3(V - I)$ for $0 < V - I < 5$. The statistical errors in the photometric measurements and the (usually-larger) flat-field uncertainties ($\lesssim 0.1$ mag) are too small to affect our conclusions. The list of stellar candidates may include faint compact galaxies as well as stars. However, we are confident that we have not excluded any stars.

We transformed the $V'$, $I'$ system to standard Johnson/Cousins $V$ and $I$ bandpasses as follows. We convolved the spectrum of each of 175 Gunn & Stryker (1983) standard stars with $V$ and $I$ throughput curves obtained from D. Schneider (private communication 1994) and with the $V'$ and $I'$ combined filter and chip response curves (Burrows et al. 1993). The $V'$ and $I'$ response curves, which differ slightly from chip to chip, were made available to us in electronic form by the Space Telescope Science Institute. We established the $V$ and $I$ zero points using the preliminary ground-based calibration by J. Holtzman (private communication 1994). These zero points agree to within $\lesssim 0.1$ mag with the values expected based on the *HST* WFC2 manual (Burrows et al. 1993). We fit the Gunn & Stryker stars to a piece-wise linear relation to obtain,

$$V_n = \alpha_V - 2.5\log(C_{V'}) + \beta_V(V-I) + \delta_n,$$
$$I_n = \alpha_I - 2.5\log(C_{I'}) + \beta_I(V-I) + \delta_n,$$
(2.1)

for standard Johnson/Cousins $V$ and $I$ on chips $n = 2, 3, 4$ as functions of the $C_{V'}$ and $C_{I'}$, the fluxes in electrons per second. We find $\alpha_V = 24.90$, $\beta_V = 0.284$, $\alpha_I = 23.79$, $\beta_I = -0.033$ for $0 < V - I < 1.90$; $\alpha_V = 25.20$, $\beta_V = 0.126$, $\alpha_I = 23.72$, $\beta_I = 0.004$ for $1.90 < V - I < 3.90$; $\alpha_V = 24.14$, $\beta_V = 0.400$, $\alpha_I = 23.17$, $\beta_I = 0.142$ for $3.90 < V - I < 5.0$; and $\delta_2 = -0.03$, $\delta_3 = 0.09$, $\delta_4 = -0.06$.

Figure 1 shows the color-magnitude diagram of stellar objects in the field; the reddest object has $V - I = 2.9$.

The star counts for the disk and thick disk were predicted using the color-magnitude relation $M_I = 3.45 + 2.25(V - I)$ for $2.0 < V - I < 4.7$. For spheroid and halo stars, we use $M_I = 7.25 + 1.4(V - I)$ for $2.0 < V - I < 4.5$. The disk and thick relation is obtained from data for dwarf stars given by Monet et al. (1992). Conservatively, the color-magnitude diagram for the spheroid and halo stars was estimated from the lower envelope of the subdwarfs.

## 3. Comparison With Dynamical Limits On Dark Matter

If the dark halo of the Galaxy were made of main-sequence stars $M_I = 14$, they would be visible in the *HST* field to a distance $d = 1.8\,\mathrm{kpc}$. We would expect $N = 65$ such stars in the field (assuming a local halo density of $9 \times 10^{-3} M_\odot\,\mathrm{pc}^{-3}$ Bahcall et al. 1983). No stars are observed with $V - I > 3.0$ and only 5 stars ($\sim 4$ of which are probably in the disk or thick disk based upon their inferred distance moduli) are detected with $2 < V - I < 3$. If there are halo stars at the main-sequence limit, $M_I = 14$, they constitute $< 6\%$ of the dark halo with 98% confidence. Making the very conservative assumption that all of the observed stars with $V > 2$ are in the halo, we find that stars in the range $10 < M_I < 13$ contribute $< 5\%$ of the dark halo and stars with $10 < M_I < 14$ contribute $< 6\%$.

The same limits on the fractional contributions to the mass of the Galaxy as were given above for the halo also apply to spheroid stars.

The maximum local column density of mass in the disk that is consistent with the observed rotation curve is $\Sigma_{\mathrm{max}} \sim 135\,M_\odot\,\mathrm{pc}^{-2}$, while the observed density of visible matter, $\Sigma_{\mathrm{obs}} = 48\,M_\odot\,\mathrm{pc}^{-2}$ (Bahcall 1984). Various studies of the vertical motions of tracer stars further constrain the $2\,\sigma$ upper limit on the total mass (disk + halo) within $\sim 1\,\mathrm{kpc}$ of the plane to $70 M_\odot\,\mathrm{pc}^{-2}$ to $85 M_\odot\,\mathrm{pc}^{-2}$, depending on the technique used (Kuijken & Gilmore 1991; Bahcall, Flynn, & Gould 1992; and Flynn & Fuchs 1994). Here, we place independent limits on the contribution to the total column density by faint red stars, of mass $m$ and absolute magnitude $M_I$, in a thin disk or a thick disk.

For an exponential disk of scale height $h$, the total number of stars counted, $N$, per unit area, $\Omega$, toward Galactic latitude $b$, is

$$\frac{dN}{d\Omega} = \frac{\Sigma}{2hm} \int_0^d dr\, r^2 \exp\left(-\frac{r \sin b}{h}\right) = \frac{\Sigma h^2 \csc^3 b}{m} \zeta\left(\frac{d \sin b}{h}\right), \qquad (3.1)$$

where $\zeta(x) = 1 - (1 + x + x^2/2)\exp(-x)$ and $d = \text{dex}(0.2[I_{\lim} - M_I] + 1)\,\text{pc}$. Note that $\zeta(3) \sim 0.6$ and $\zeta(x) \sim 1$ for $x \gtrsim 5$ (luminous stars). For an ensemble of stars, we may replace the quantity $\zeta/m$ by its average, $\zeta/m \to \langle \zeta/m \rangle$.

For $V - I > 3.0$, we detect $N_{\text{obs}} = 0$ stars. We can therefore rule out $N \geq 4$ at the 98% confidence level and hence place an upper limit of

$$\Sigma < 8\, M_\odot\, \text{pc}^{-2} \left(\left\langle \frac{\zeta}{m} \right\rangle 0.2\, M_\odot\right)^{-1}. \qquad (3.2)$$

This corresponds to a limit of $\Sigma < 8\, M_\odot\, \text{pc}^{-2}$ for stars with $10.2 < M_I < 13.5$ where $\zeta \sim 1$. This would correspond to $< 15\%$ of the total mass of the disk. If all the stars have $M_I = 14$, the limit actually becomes stronger since the stars have lower mass. Taking account of this and the lower $\zeta$, we find $\Sigma < 5\, M_\odot\, \text{pc}^{-2}$ (for $M_I = 14$) and $\Sigma < 8\, M_\odot\, \text{pc}^{-2}$ (for $M_I = 15$). Previous estimates gave $\Sigma \sim 16\, M_\odot\, \text{pc}^{-2}$ for all thin disk stars with $M_V \geq 6.8$, i.e., $V - I \geq 1.1$ (Bahcall 1984).

For a thick disk of scale height $h \sim 1.4\,\text{kpc}$ (Gilmore & Reid 1983), we obtain 98% confidence limits of $\Sigma < 3.3\,M_\odot\,\text{pc}^{-2}$ for stars with $V - I > 3.0$ ($M_I > 10.2$) and brighter than $M_I < 14$ and $\Sigma < 10\,M_\odot\,\text{pc}^{-2}$ for $M_I < 15$.

## 4. Constraints On Disk and Spheroid Luminosity Functions

The faint end of the disk and spheroid luminosity functions (LFs) have in the past been measured primarily from nearby stars. Important information about the LFs has been obtained from star counts down to $V \sim 20$, beyond which star/galaxy confusion makes further progress difficult with ground-based images. However, nearby samples suffer from several limitations. For the spheroid, the absolute number of faint nearby stars is small. Moreover, kinematically selected spheroid samples may be biased. For the disk, a much larger fraction of nearby stars are young than is characteristic of the disk population as a whole. For the faintest stars, the volume probed by the *HST* WFP2 field is equivalent to a $4\pi$ all sky sample complete to $I = 13$, much deeper than any local volume survey. Star counts from this field therefore provide the best constraints yet on the shape of the faint end of the LFs.

We establish constraints by comparing the predicted number of stars, $N$, in the region $V - I > 3.0$, $I < 25.3$ with the observed counts, $N_{\text{obs}} = 0$. We parameterize the faint end by a power-law exponent, $\gamma$, with the cumulative number of stars, $\Phi$, growing as a function of $V$-band luminosity, $d\Phi/d\ln L_V \propto L_V^{-\gamma}$.

For the disk, we find that a conventional Bahcall-Soneira model (Bahcall 1986) with a flat ($\gamma = 0$) LF beyond $M_V = 15$ (where the LF was previously poorly constrained) predicts 4.9 stars and is thus ruled out with 99% confidence. The disk LF therefore falls for $15 > M_V > 18.7$, i.e., to the hydrogen burning limit.

For the spheroid, the LF has been poorly constrained beyond $M_V = 6$. Normalizing the spheroid LF locally to 1/500 (Bahcall, Schmidt, and Soneira 1983) of the disk LF at $M_V = 6$, we find that power laws with $\gamma > 0.32$ are ruled out at the 98% confidence level. If the spheroid is normalized locally at 1/800 of the disk LF (as suggested by Gilmore & Reid 1983), then $\gamma > 0.37$ is similarly ruled out. The *HST* field effectively probes the spheroid LF over a mag range $14 \lesssim M_V \lesssim 17$.

The upper limits obtained here disagree with the average value, $\gamma = 0.63$, found by Richer & Fahlman (1992) from faint star counts over the interval $6 < M_V < 14$. Richer & Fahlman argue that their result showed that the spheroid LF is consistent with that found in some globular clusters. However, using their value of $\gamma = 0.63$ and a local normalization of 1/500 at $M_V = 6$, we calculate we should have detected $N \sim 75$ stars with $V - I > 3.0$. in the *HST* field, against none observed. We should also have detected $N \sim 45$ with $2 < V - I < 3$ ($12 < M_V < 14.5$) against $N_{\rm obs} \leq 5$ observed. The large slope obtained by Richer & Fahlman may result from the fact that their normalization of the spheroid at $M_V = 5\text{-}6$ is $10^{-8}$ stars mag$^{-1}$ pc$^{-3}$, two orders of magnitude below the measured value (see Bahcall, Schmidt, & Soneira 1983). Also, if the local spheroid denisty is as large as proposed by Caldwell & Ostriker (1981), $1.1 \times 10^{-3} \, M_\odot$ pc$^{-3}$, then this mass must be primarily in objects fainter than $M_V \sim 16$.

## 5. Microlensing and Star Counts

Two groups have recently detected a total of 6 candidate microlensing events toward the LMC. (Alcock et al. 1993, MACHO; Alcock et al. 1994; Aubourg et al. 1993, EROS). These events were found, as originally suggested by Paczyński (1986), by monitoring more than 10 million LMC stars over about 300 days, in order to detect massive compact halo objects (Machos) that may make up the dark matter in galactic halos.

On the basis of the 4 detections by MACHO, we estimate a minimum optical depth of $\tau_{\min} = 4 \times 10^{-8}$ for events long enough to have been observed a number of times (and assuming a $f = 100\%$ detection efficiency and that all candidates are true microlenses). The true optical depth is probably higher because $f < 100\%$. Several workers have suggested that the events are due to a faint, previously unseen, component of the disk, thick disk, or spheroid (Gould, Miralda-Escudé, & Bahcall 1994; Gould 1994a; Giudice, Mollerach, & Roulet 1994). Here we show that low-mass stars (late M dwarfs or subdwarfs) do not contribute significantly to the observed microlensing events.

Consider a population of stars that are distributed exponentially in height above the Galactic plane with scale height $h$ and with total column density $\Sigma$. From equation (3.1) and Gould (1994a), we find the number of star counts in an $\Omega = 4.4$ square arcmin field required to account for a given the optical depth, $\tau$, toward the LMC is

$$N = \frac{h \langle \zeta/m \rangle c^2}{2\pi G} \frac{\sin^2 b_{\rm LMC}}{|\sin^3 b|} \Omega \tau = 54 \left(\frac{h}{350\,{\rm pc}}\right) \left(\left\langle \frac{\zeta}{m} \right\rangle 0.2 M_\odot\right) \left(\frac{\tau}{\tau_{\min}}\right), \quad (5.1)$$

where for the sake of definiteness we have written the result in terms of $\tau_{\min} =$

$4 \times 10^{-8}$. Because the LMC and the *HST* field are both at relatively high latitude, we have ignored the radial dependence of the disk density.

Equation (5.1), combined with the fact that we detect no stars with $V - I > 3.0$, implies that thin-disk red stars $M_I < 13.5$, bright enough to be seen at a distance corresponding to five disk scale heights $d = 5h \csc b = 2.3 \,\mathrm{kpc}$ are responsible for $< 0.08 \, \tau_{\min}$, with 98% confidence. If all the stars have $M_I = 14$, the maximum contribution is smaller, $< 0.05 \, \tau_{\min}$ ($< 0.08 \, \tau_{\min}$ at $M_I = 15$).

For a thick disk of scale height $h \sim 1.4 \,\mathrm{kpc}$ (as suggested by Gilmore & Reid 1983), red stars with $M_I < 14$ contribute $< 0.12 \tau_{\min}$. The limit rises to $< 0.36 \tau_{\min}$ for $M_I = 15$.

Gould (1994b) has calculated the optical depth to a source lying within a spheroidal mass distribution with radial profile $r^{-n}$ as seen by an infinitely distant observer. These same formulae apply for an observer within a spheroidal mass distribution and looking at an infinitely distant source. Since the LMC is at a finite distance, there is a $\sim 20\%$ reduction in $\tau$.

The number of stars of mass $m$ and absolute magnitude $M_I$ observed per unit area toward Galactic coordinates $l$ and $b$ is

$$\frac{dN}{d\Omega} = \frac{\rho_0 R_0^n}{m} \int_0^{d(M_I)} dr \frac{r^2}{(r^2 + R^2 - 2rR_0 \cos \ell \cos b)^{n/2}}. \tag{5.2}$$

The expected number of stars in the *HST* field per unit optical depth toward the LMC rises steeply with the index, $n$. We investigate the $n = 3$ model in order to place limits on all steeper spheroid models. For a poplulation of spheroid stars with $V - I = 3.0$ ($M_I = 11.5$) we expect to detect $N = 70\tau/\tau_{\min}$ stars versus $N_{\text{obs}} \leq 1$ detected with $V - I > 2.6$. At $V - I = 2.5$ and $V - I = 2.0$, the number of expected stars rises to $N = 140\tau/\tau_{\min}$ and $N = 270\tau/\tau_{\min}$ respectively, while the total number of stars detected rises to $N_{\text{obs}} \leq 5$. These calculations imply that spheroid stars $10 < M_I < 12$ account for $< 0.09\tau_{\min}$ with 98% confidence. Stars with $12 < M_I < 13.5$ account for $< 1.0\tau_{\min}$ at the 98% confidence level.

## 6. Summary and Discussion

Faint red stars down to the hydrogen-burning limit do not contribute significantly to the mass budget of the Milky Way. Faint red stars in the dark halo or in the spheroid constitute $< 6\%$ of the mass of the Galaxy. M-dwarfs in the disk constitute less than $< 15\%$ of the disk mass.

Late M dwarfs in the disk or in a thick disk also do not contribute significantly to the microlensing events that have been detected toward the LMC. If main sequence stars in the spheroid are responsible for these microlensing events, they must be fainter than $M_I = 13$.

**Acknowledgements**: This research was supported in part by NASA grant #NAG5-1618. The work is based primarily on observations with the NASA/ESA *Hubble Space Telescope*, obtained at the Space Telescope Science Institute, which is operated by the Association of Universities for Research in Astronomy, Inc., under NASA contract NAS5-26555.

FIGURE CAPTIONS

1) Color-magnitude diagram of stellar objects detected in a 4.4 square arcmin field toward $l = 241$ $b = -51$, using the repaired *HST* WFC2. Standard Johnson/Cousins $V$ and $I$ magnitudes were obtained from the instrumental magnitudes using equation (2.1). The mean magnitude $I$-band limit for the 3 chips (*solid line*) was evaluated by subsituting the flux limits (see text) into equation (2.1) with $\delta_n = 0$. Objects above the flux limit in either $V$ or $I$ are included in the study.

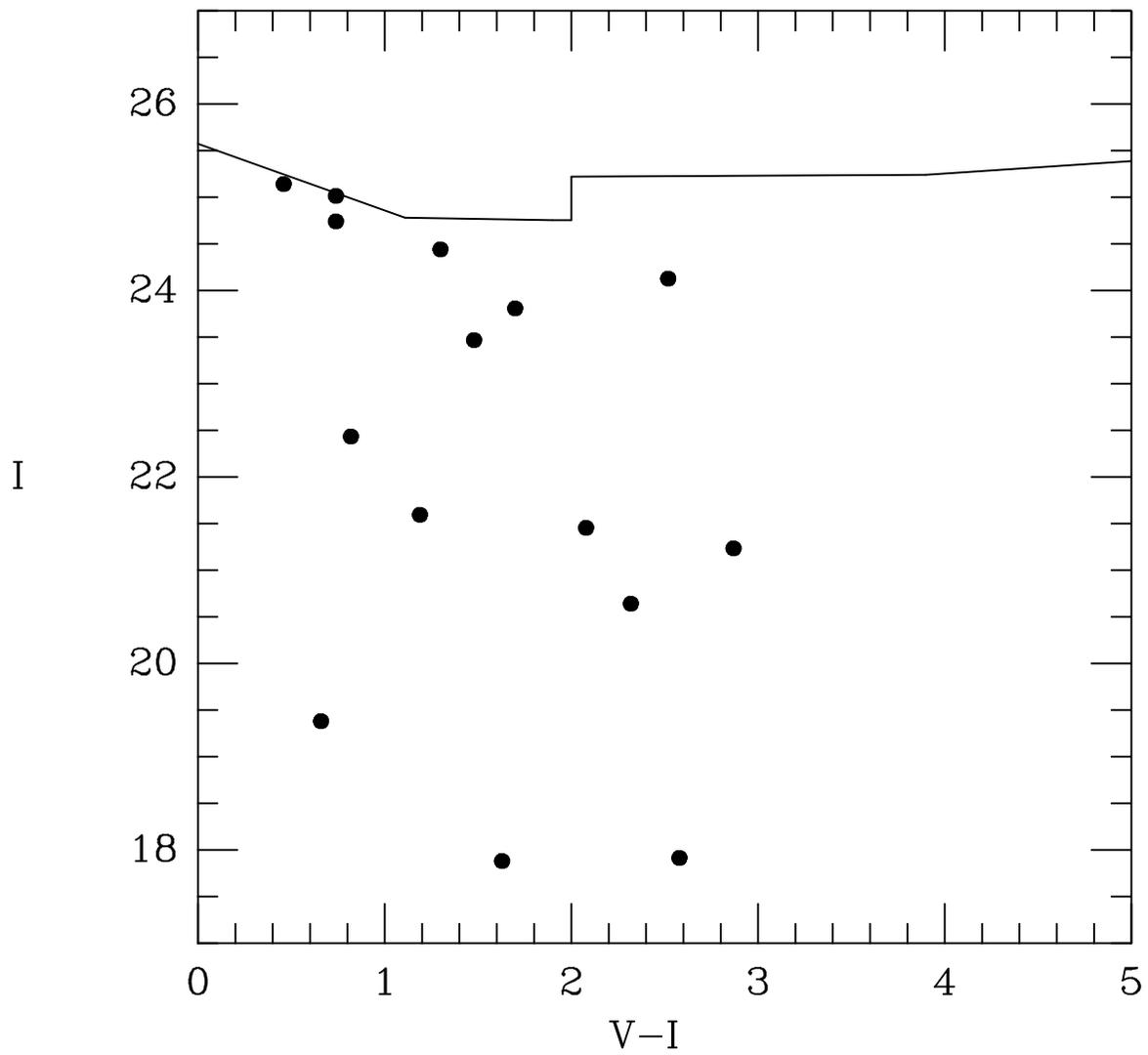